\documentclass[12pt]{iopart}

\usepackage{graphicx}

\newcommand\mb[1]{\mathbf{#1}}
\begin{document}


\title{Molecular Dynamics Simulation of Vascular Network Formation }

\author{Paolo Butt\`a$^1$, Fiammetta Cerreti$^1$,
Vito~D~P~Servedio$^{2,3}$ and Livio Triolo$^4$}%
\address{$^1$ Dipartimento di Matematica, Universit\`a di Roma ``La
Sapienza'', P.le A. Moro 2, 00185 Roma, Italy}%
\address{$^2 $Dipartimento di Fisica, Universit\`{a} di Roma ``La
Sapienza'', P.le A. Moro 2, 00185 Roma, Italy}
\address{$^3$ Museo Storico della Fisica e Centro Studi e Ricerche ``Enrico Fermi'',
Compendio Viminale, 00184 Roma, Italy}
\address{$^4$ Dipartimento di Matematica, 
Universit\`a di Roma ``Tor Vergata'', Via della Ricerca Scientifica
1, 00133 Roma, Italy}%
\ead{\mailto{butta@mat.uniroma1.it}, \mailto{cerreti@mat.uniroma1.it},
\mailto{vito.servedio@roma1.infn.it}, \mailto{triolo@mat.uniroma2.it}}


\begin{abstract}
\noindent 
Endothelial cells are responsible for the formation of the capillary
blood vessel network. We describe a system of endothelial cells by
means of two-dimensional molecular dynamics simulations of point-like
particles.  Cells' motion is governed by the gradient of the
concentration of a chemical substance that they produce
(chemotaxis). The typical time of degradation of the chemical
substance introduces a characteristic length in the system. We show
that point-like model cells form network resembling structures 
tuned by this
characteristic length, before collapsing altogether.  Successively,
we improve the non-realistic point-like model cells by introducing an
isotropic strong repulsive force between them and a velocity dependent
force mimicking the observed peculiarity of endothelial cells to
preserve the direction of their motion (persistence).  This more
realistic model does not show a clear network formation.  We ascribe
this partial fault in reproducing the experiments to the static
geometry of our model cells that, in reality, change their shapes by
elongating toward neighboring cells.
\end{abstract}


\maketitle

\section{\label{sec:intro}Introduction}

\noindent Vascular networks are complex structures resulting from
the interaction and self organization of endothelial cells (ECs).
Their formation is a fundamental process occurring in embryonic
development and in tumor vascularization. In order to optimize the
function of providing oxygen to tissues, vascular network
topological structure has to involve a characteristic length
practically dictated by the diffusion coefficient of oxygen
\cite{ambrosi05}. In fact, observation of these networks reveals that
they consist of a collection of nodes connected by thin chords with
approximately the same length.  In this work, we study the process of
formation of vascular networks by means of two-dimensional off-lattice
molecular dynamics simulations involving a finite number of
interacting simple units, modeling endothelial cells.  The interaction
among cells is due to the presence of a chemical signal, in turn
produced by the cells themselves. This mechanism of motion is called
\emph{chemotaxis}, a mechanism still object of intensive experimental
and theoretical research.  The relevance of chemotaxis reflects its
important role in many situations of biomedical interest such as wound
healing, embryonic development, vascularization, angiogenesis, cell
aggregation, to cite a few.
On the theoretical side, the first partial differential equation based
models of chemotaxis appeared in the early '70s~\cite{kelseg}, and
were soon followed by ``more microscopic'' models (eg.\ the Reinforced
Random Walk) and kinetic models~\cite{othmstev}.

The study of the particular biological process of vascular network formation and
its relations to tumor vascularization has also been accomplished by means
of several mathematical models.  First studies were presented by
Murray et al.\ \cite{MO,MLVM}, who explained the phenomenon by
focusing mainly on its mechanical aspects, i.e.\ on the interaction
between cells and the substrate.  Gamba et al.\ %
\cite{Tor1,Tor2,Tor3,Tor7} proposed a continuous model, based on
chemotaxis, which applies to early stages of \emph{in vitro}
vasculogenesis, performed with Human Umbilical-Vein ECs (HUVEC)
cultured on a gel matrix.  More recently, some of these authors
managed to unify both the mechanical aspect and the chemical one into
a more complete model
\cite{Man,Tor5}.

Blood vessel formation may be divided into two different processes. In
the first stage, occurring in embryonic development, ECs organize into
a primitive vascular network (\emph{Vasculogenesis}).  In a second
moment, existing vessels split and remodel in order to extend the
circulation of blood into previously avascular regions by a mechanism
of controlled migration and proliferation of the ECs
(\emph{Angiogenesis}) \cite{PS}.  ECs are the most essential component
of the vessel network: each vessel, from the largest one to the
smallest one, is composed by a monolayer of ECs (called
\emph{endothelium}), arranged in a mosaic pattern around a central
lumen, into which blood flows.  In the capillaries the endothelium may
even consist of just a single EC, rolled up on itself to form the
lumen.

Although there are several mechanisms involved during vessel
formation, in this work we shall focus on the characteristic migration
motion of cells driven in response to an external chemical stimulus:
the chemotaxis. ECs secrete an attractive chemical factor, the
Vascular Growth Factor-A (VEGF-A), while they start to migrate. Each
of them perceives the chemical signal with its receptors at its
extremities and starts to move along the chemical concentration field
gradient, toward areas of higher concentration corresponding to higher
density of cells.  ECs are able to move extending tiny protrusions, the
{\itshape pseudopodia}, on the side of the higher concentration. The
pseudopodia attach to the substratum, via adhesion molecules, and
pull the cell in that direction.

In parallel with chemotaxis, another mechanism resulting in cell
motion is the \emph{haptotaxis}, i.e.\ the movement of cells along an
adhesive gradient: the substratum is not usually homogeneous and its
varying density can affect cell adhesion and hence migration. We will
not consider haptotaxis in this work.  Cell-cell and cell-membrane
contacts are really essential in the process of vascular network
formation and their loss can cause cell apoptosis (death)
\cite{PS,DLFL,LS,Turner,Tor8,Tor9}.

\section{Review of the experimental data}
\noindent 
The experimental data on which all theoretical studies till now
hinge, are those collected by tracking the behavior of cells
initially displaced at random onto a proteic gel matrix, generically
called  Extra-Cellular Matrix, which mimics at best the
original living environment.
In our analysis, we explicitly refer to the \emph{in
vitro} vasculogenesis experiments of Gamba et al.\ \cite{Tor1}, and
shall use the same numerical values of parameters therein introduced.
In the experiments, HUVEC cells are randomly dispersed and cultured
on a gel matrix (\emph{Matrigel}) of linear size $l=1$,2,4,8~mm. Cells
sediment by gravity on the matrix and then move on its horizontal
surface, showing the ability of self-organizing in a structured
network characterized by a natural length scale. The whole process
takes about $T=12$ hours.\\
Four fundamental steps can be distinguished in the experiments
\cite{Tor6}:
\begin{itemize}
\item[$\cdot$]
During the first two hours, the ECs
start to move by choosing a particular direction dictated by the
gradient of the concentration of the chemical substance VEGF-A.
Single ECs migrate until collision with neighboring cells, keeping a
practically fixed direction with a small superimposed random velocity
component.  The peculiarity of ECs of maintaining the same direction
of motion is known as \emph{persistence}, and has been explained by
cells inertia in rearranging their shapes.  In fact, in order to
change direction of motion, ECs have first to elongate towards the new
direction, with the result that to change path is a relatively slow
process.  In this phase of amoeboid motion, the
mechanical interactions with the substrate are weak.
\item[$\cdot$]
After collision, ECs attach to their neighboring cells eventually
forming a continuous multicellular network.  They assume a more
elongated shape and multiply the number of adhesion sites.  In this
phase the motion is slower than in the previous step.
\item[$\cdot$]
In the third phase the mechanical interactions become essential as the
network slowly moves, undergoing a thinning process that would leave
the overall structure mainly unalterated.
\item[$\cdot$]
Finally cells fold up to create the lumen.
\end{itemize}
The final capillary-like network can be represented as a collection
of nodes connected by chords, whose experimentally measured average
length stays around 200~$\mu$m for values of the cell density
between 100 to 200~cells/mm$^2$. Outside this range no network
develops. More precisely, below a critical value of 100~cells/mm$^2$,
groups of disconnected structures form, while at higher densities
(above 200-300~cells/mm$^2$) the mean chord thickness grows to hold
an increasing number of cells and the structure resembles a
continuous carpet with holes (\textit{swiss cheese} pattern).

\section{Theoretical model}
\noindent In this article we present an off-lattice particle model of
vasculogenesis where the equations of motion are governed by the
gradient of the concentration of a chemoattractant substance
produced by the particles themselves.  The discrete $N$-particle
system we are proposing, gives evidence of the important role of the
pure chemotaxis process in forming well structured networks with a
characteristic chord length size.
\begin{figure*}
\begin{center}
\begin{tabular}{cccc}
        A&B&C&D\\
        \includegraphics[width=.24\textwidth,keepaspectratio]{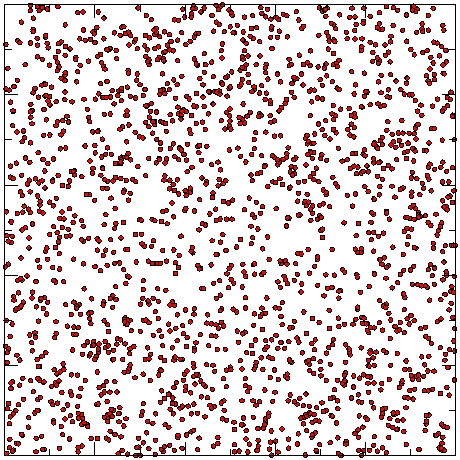}&
        \includegraphics[width=.24\textwidth,keepaspectratio]{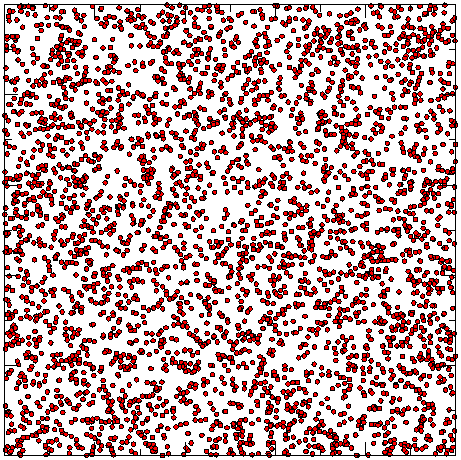}&
        \includegraphics[width=.24\textwidth,keepaspectratio]{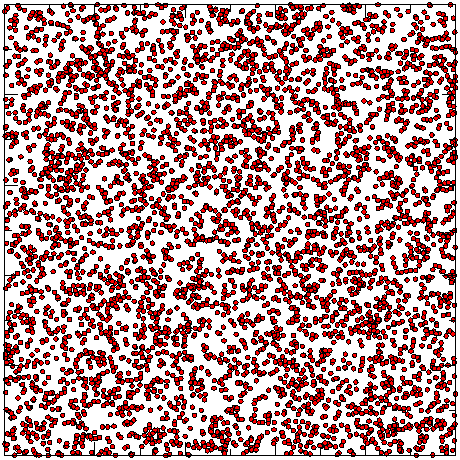}&
        \includegraphics[width=.24\textwidth,keepaspectratio]{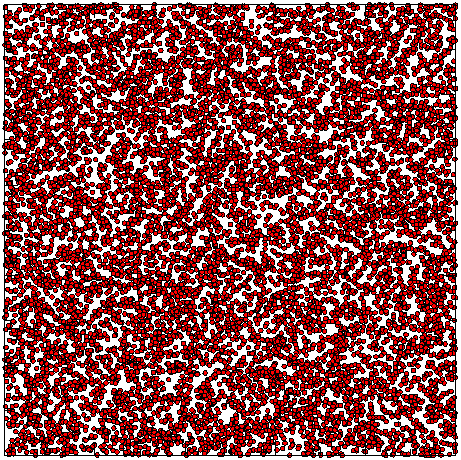}\\
        \includegraphics[width=.24\textwidth,keepaspectratio]{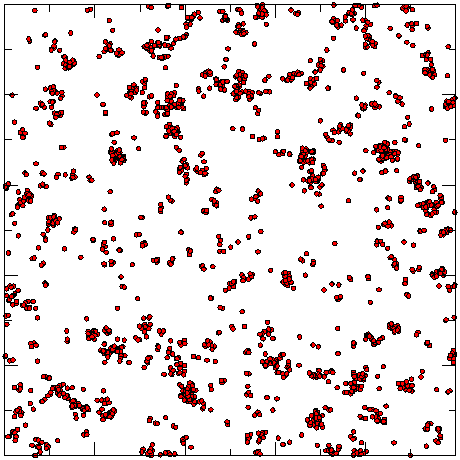}&
        \includegraphics[width=.24\textwidth,keepaspectratio]{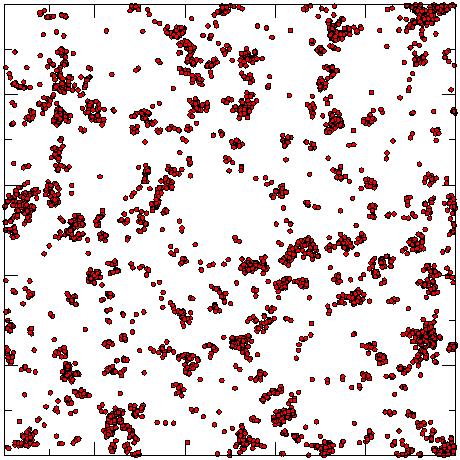}&
        \includegraphics[width=.24\textwidth,keepaspectratio]{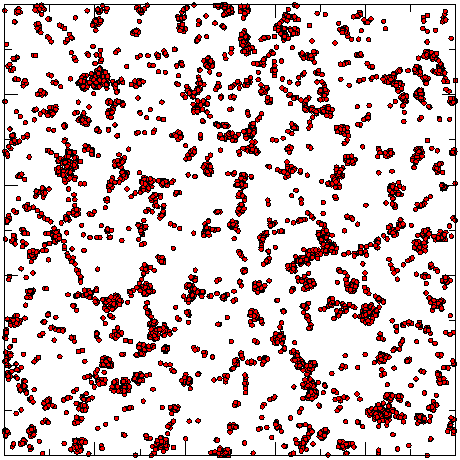}&
        \includegraphics[width=.24\textwidth,keepaspectratio]{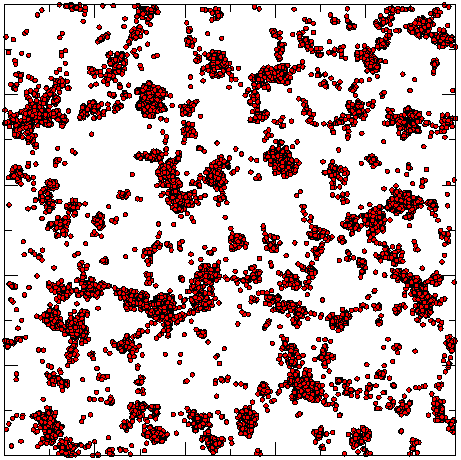}\\
\end{tabular}
\caption{\label{fig:nohc} Results of the simulations  of
point-like particle under the sole effect of chemotaxis inside the box
with edge 1~mm$=5\ell$; $\ell=\sqrt{D\tau}$ indicates the
characteristic length of the process. Each column involves different
particle numbers: (A) 2000 cells, (B) 3750 cells, (C) 5000 cells, (D)
11250 cells. The top row shows the initial random particle
displacement. The bottom row shows the systems after the dynamics
produced a network-like resembling structure. Since this structure
appears during a brief transient before the expected structural
collapse into a single agglomerate, the simulation times of these
snapshots were chosen qualitatively after visual inspection and
roughly correspond to $T=2$ hours of laboratory time (much less than
experimental times due to the large cell densities resulting in
unrealistically large chemo-attractant concentration gradients).
In these simulations we used the dimensionless $\alpha=1$.}
\end{center}
\end{figure*}

We developed our model with increasing complexity, refining it by
gradually adding features that would allow a closer resemblance with
experiments.  Particles, which we shall also refer to as ``cells'' in
the following, are constrained inside a square box of given edge $L$
with periodic boundary conditions. The number of cells will be kept
constant during the simulations, i.e.\ we will consider neither cell
creation nor cell destruction.\\
At first, we consider cells as adimensional point-like particles
moving only under the effect of the concentration gradient of the
chemoattractant substance $\nabla c(\mb{x},t)$.  The chemoattractant
is released by cells. It diffuses according to a diffusion
coefficient $D\approx 10\ \mu$m$^2$/s and degrades within a
characteristic finite time $\tau\approx 64$~min. The combination of
the diffusion and degradation processes introduces a characteristic
length $\ell=\sqrt{D\tau}$ in the system, with $\ell\approx 200\
\mu$m. As a further step, we introduced in the system a dynamical
friction proportional to the velocity of cells in order to simulate
the dragging force to which cells are subjected in the matrigel. The
net result of this friction is to slow down the overall simulation
time, leaving unaltered the main features of the system history.
Although point-like cells are a very simplified representation of real
cells, their motion, driven by chemotaxis, yields the formation of the
expected network of filaments with a characteristic length $\ell$
(see bottom row of Fig.~\ref{fig:nohc}).

In order to further refine the simulation, we introduced an inelastic
isotropic repulsion mechanism between cells (see the following
paragraph for its definition), imitating the fact that cells do not
penetrate each other in reality. Since the chemotactic field acts as
an attractive potential between cells, the result of both processes is
to stick together cells after collision.  Cells are no more
adimensional but now possess their own effective radius 
$r\approx 10\ \mu$m. 
The introduction of a cell effective radius changes sensibly
the simulation. One important side effect is that of imposing a limit
on the density of cells in the simulation, which must lie well below
the close-packing density.
\begin{figure*}
\begin{center}
\begin{tabular}{cccc}
        A&B&C&D\\
        \includegraphics[width=.24\textwidth,keepaspectratio]{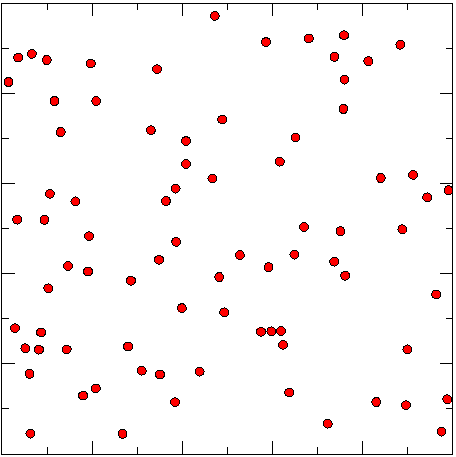}&
        \includegraphics[width=.24\textwidth,keepaspectratio]{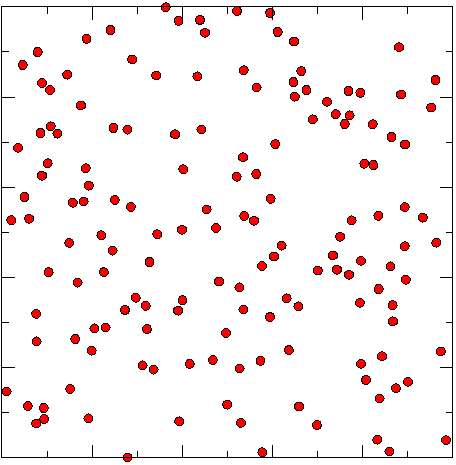}&
        \includegraphics[width=.24\textwidth,keepaspectratio]{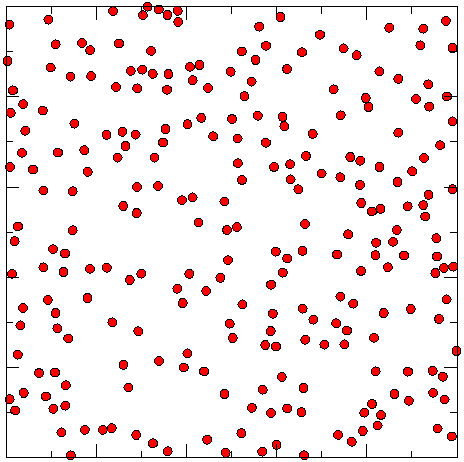}&
        \includegraphics[width=.24\textwidth,keepaspectratio]{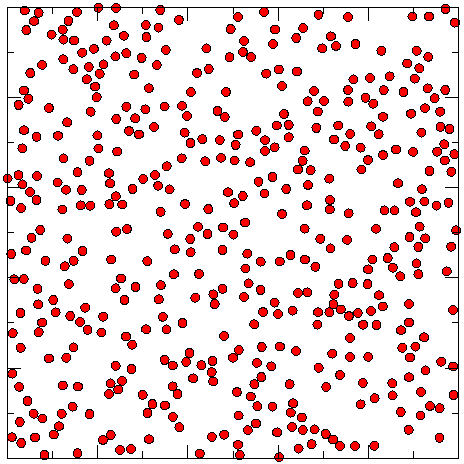}\\
        \includegraphics[width=.24\textwidth,keepaspectratio]{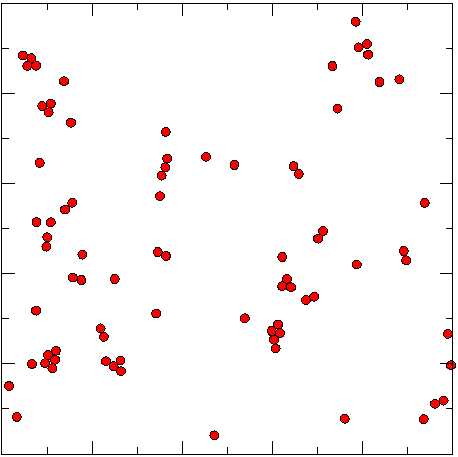}&
        \includegraphics[width=.24\textwidth,keepaspectratio]{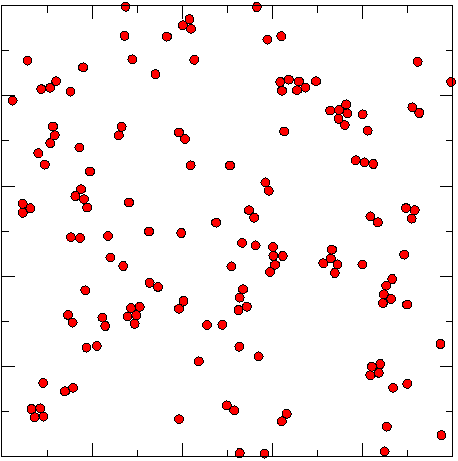}&
        \includegraphics[width=.24\textwidth,keepaspectratio]{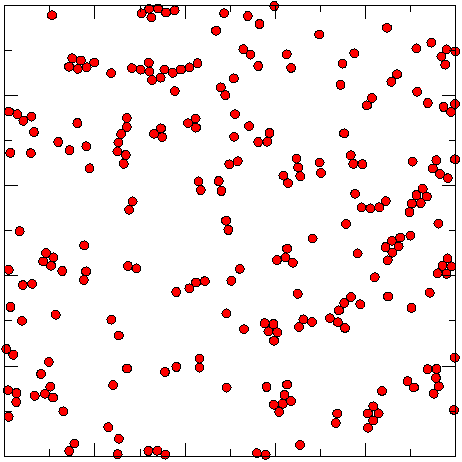}&
        \includegraphics[width=.24\textwidth,keepaspectratio]{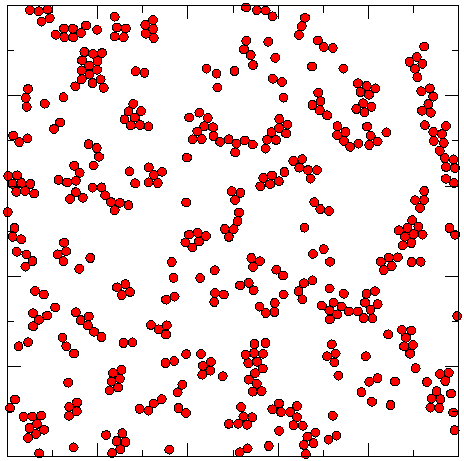}\\
\end{tabular}
\caption{\label{fig:hctr}
Simulation results of particles in presence of an inelastic repulsion
potential between cells and a persistence force together with
chemotaxis, inside the box with edge 1~mm$=5\ell$; $\ell=\sqrt{D\tau}$
indicates the characteristic length of the process.  Each column
involves different particle numbers: (A) 80 cells, (B) 150 cells, (C)
250 cells, (D) 450 cells. The first row shows the initial random
particle displacement. The second row shows the systems after the
dynamics produced a possible network-like structure. The simulation
times at which the networks form were chosen qualitatively after
visual inspection, as in Fig.~\ref{fig:nohc}.  The chosen final
simulation time corresponds to about $T\simeq 13$ hours, i.e.\ a value
comparable with the observed network formation times of experiments.
The diameter of cells is of $20\mu m=0.1\ell$ and is faithfully
represented in scale. 
In these simulations we fixed the dimensionless
constants to $\alpha=1$, $\gamma=1$, $\kappa=1000$, $\nu=1$, $\eta=100$.}
\end{center}
\end{figure*}
The last refinement deals with the problem of the ``cell persistence''
of motion, i.e.\ the observed large inertia of cells in changing the
direction of their motion.  The solution to this issue in terms of
pure galilean inertia proposed by Gamba et al.\ in their continuous
model \cite{Tor1} has been criticized in \cite{MG2}, where it was
pointed out that cells must rely onto a more involved mechanism of
resistance to changes of direction.  As a result some of the authors
of \cite{Tor1} recently proposed variations of their model to
explicitly include cell persistence \cite{Tor5,Tor6}.  In our case, we
use the advantage of a molecular dynamic simulation to have full
control of all forces acting on all cells at each time step.  For each
cell, we simply reduce the component of the gradient of the chemical
field along the direction orthogonal to the direction of cell motion
by a factor
$\beta_i=\kappa|\mb{v}_i|/(\kappa|\mb{v}_i|+|\mb{\nabla}c(\mb x_i)|)$
with $\kappa$ a constant modulating the effect of persistence and
$(\mb x_i,\mb{v}_i)$ the position and velocity of the $i$-th cell.

To summarize, the dynamical system of equations we solve with $i=1
\ldots N$ is
\begin{eqnarray*}
    \dot{\mb x}_i(t)&=&\mb v_i(t)\\
    \dot{\mb v}_i(t)&=&
        \mu\nabla c(\mb x_i(t),t) +
        \mb F_{\mathrm{IR}} + \mb F_\mathrm{T} + \mb F_\mathrm{F}\\
    \partial_t c(\mb x,t)&=&
    D \Delta c(\mb x,t)-\frac{c(\mb x,t)}{\tau}+
    \alpha \sum_{j=1}^NJ(\mb x-\mb x_j(t)),
\end{eqnarray*}
where $D$, $\alpha$, and $\tau$ are respectively the diffusion
coefficient, the rate of production and the characteristic
degradation time of the soluble chemo-attractant mediator, and $\mu$
measures the strength of the cell response to the chemical factor.
Here $c(\mb x,t)$ is the total chemical field acting on the position
$\mb x$ at time $t$. By linearity it is given by the sum of the
chemotactic fields $c_j(\mb x,t)$ generated by all cells
$\sum_{j=1}^N\,c_j(\mb x,t)$.\\
The force $\mb F_{\mathrm{IR}}$
stands for the inelastic isotropic repulsion force between two
colliding cells. 
Let $\mb d$ be the vector joining the centers cells $i$ and $j$,
we define the force 
$\mb F_{\mathrm{IR}} = \nu (2r-d)\mb{\hat{d}}-\eta(\mb v_i \cdot \mb{\hat{d}})\mb{\hat{d}}$ for
$d<2r$ (overlapping cells) and zero otherwise,
where $r$ is the cell radius, $\nu$ is an elastic constant,
 $\eta$ is an inelastic coefficient and $\mb{\hat{d}}=\mb d/d$.\\
$\mb F_{\mathrm{T}}=-\beta_i\mu\nabla c_{i,\perp \mb v}$ the cell 
persistence force discussed above, with the symbol $\nabla
c_{i,\perp \mb v}$ standing for the projection of the gradient
$\nabla c(\mb x_i)$ orthogonal to the direction of $\mb v_i$.\\
$\mb F_{\mathrm{F}} = -\gamma\mb v_i$ is the non-conservative friction
term.\\
The function $J(\mb x)$ is responsible of chemo-attractant
production. It is an even non-negative function of $\mb x$,
normalized to unity in the whole space. We took $J(\mb x)$ as a step
function with constant value inside the circle $|\mb x|<r$ and zero
outside.

In order to match the experimental set up conditions, cell initial
positions were extracted at random with a poissonian process, while
all velocities were set to zero. In the case of two-dimensional cells
with repulsion, cells were dropped in the random extracted position
only if no overlap with other cells occurred, otherwise another
position had to be extracted.  The concentration of the
chemo-attractant was also initially set to zero.

We integrated the equations of motion by means of a standard Verlet
algorithm \cite{allen} with time step $4\times10^{-5}\tau\simeq
0.15$~s, 
while the numerical solution of the
diffusion partial differential equation was accomplished by using a
second order explicit finite differences scheme \cite{NRC} on a square
grid with step equal to $2.5\times10^{-2}\ell\simeq5\mu$m.
The edge $L$ of the simulation box was set to $5\ell\simeq1$~mm. 
By posing $x^*=x/\sqrt{D\tau}$, $t^*=t/\tau$ and $c^*=c\mu\tau/D$
the equations can be written in terms of the dimensionless starred unknowns, so that
all the dependence on the constant parameters $D$, $\mu$, $\tau$
is carried upon a rescaled value of $\alpha^*=\alpha\tau^2\mu/D$ and
all forces $\mb F^*_\mathrm{\{IR,T,F\}}=\mb F_\mathrm{\{IR,T,F\}}\tau^2/\sqrt{D\tau}$.

As already mentioned, we performed our simulations with
increasing complexity.  Firstly we consider point-like model cells
moving inside the chemotactic field.  Secondly, we switched on the
inelastic cell-cell repulsion term $\mb F_\mathrm{IR}$ and the
velocity dependent persistence force $\mb F_\mathrm{T}$.  In both
cases, we included the dynamical non-conservative friction force $\mb
F_\mathrm{F}$ in the simulations, the presence of which did not change
substantially the results.
\begin{figure*}
\begin{center}
\includegraphics[width=.5\textwidth,keepaspectratio]{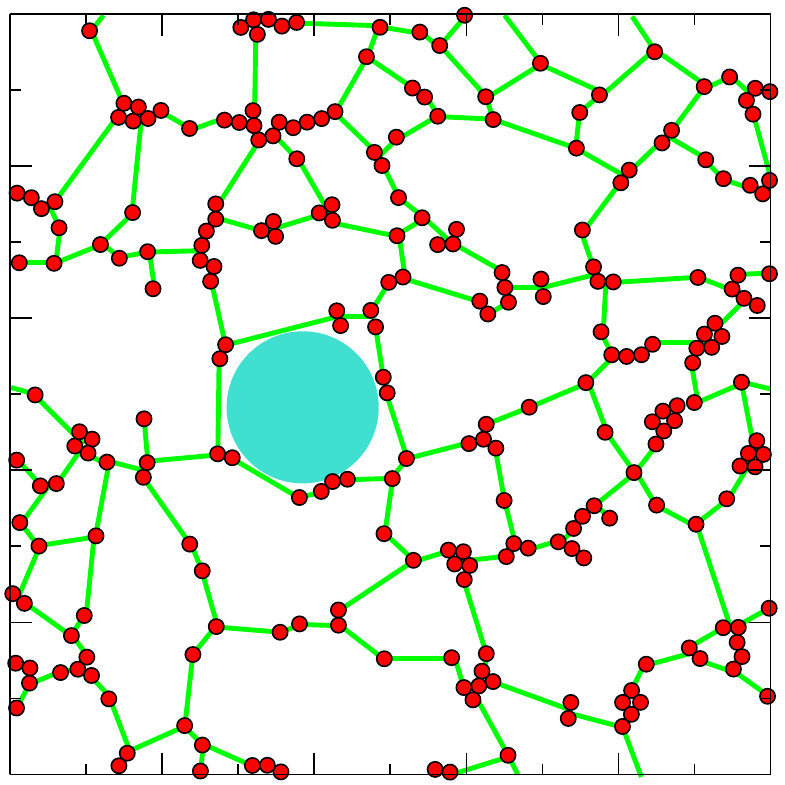}
\caption{
Same as Fig.~\ref{fig:hctr}C but here we draw by hand the possible
resulting network. In order to compare empty regions with the
characteristic area $\ell^{\,2}$ we show a circle with diameter equal
to $\ell$. 
The underlying network was drawn in a qualitative way by connecting 
nearby cells. We suggest that a more realistic model with elongating cells might
reproduce the displayed network structure.
\label{fig:net}
}
\end{center}
\end{figure*}

\section{Results}
\noindent
The results of the simulation of point-like cells under chemotaxis,
without the effect of repulsion and persistence forces, are shown in
Fig.~\ref{fig:nohc}, where the system is
analyzed depending on the number of particles.  These simulations, far
from representing the dynamics of real cell populations, are
nevertheless interesting since they deliver a picture of the
capillary network with a characteristic chord size of length
$\ell\approx 200~\mu$m and uncover the different behavior of the
system with respect to the cell density.  For $N=2000$, network chords
do not develop and after a short transient of slow motion, in which
the chemical substance diffuses throughout the simulation box,
cells collapse all together.  The situation improves as the number of
particles increases and the best network is obtained for $N=5000$. In
that case, thin filaments are visible with a thickness of very few
cells. Finally, for $N=11250$ chords are obtained with a thickness of
many cells. In some senses this last situation resembles the
\emph{swiss cheese} pattern visible in the experiments. The
characteristic length is still detectable in the form of holes with
characteristic area value around $\ell^{\,2}$.\\
We notice that in the case of pointless particles, cell densities
resulting in vascular networks are very large compared to the
experimental ones.  This is obviously an issue stemming from the
vanishing excluded volume of the chosen model particles that represent
cells.  
Moreover, these unrealistic high cellular densities result in a
correspondingly high production of chemo-actractant and consequently
high chemotactic field gradients, so that the
simulation runs faster than the characteristic times of experiments.
In fact, with these semplified model we find the formation of network
resembling structures after $T\simeq 2$~hours, much less than the
experimental value of 12~hours. 

Without both the repulsion and persistence terms, the organized
capillary network structure arises as a brief transient, after which
cells collapse all together. The appearance of network structures only
during a transient is a feature shared by other different models of
continuous type \cite{gambapc}.  At this stage, the dynamical friction
term, although strictly unnecessary, helps to lengthen the duration of
the transient. Therefore, as a first result we observe that a simple
molecular dynamics model captures the essence of the process of
chemotaxis, which is capable of organizing cells in a non-trivial
functional network displacement, although only during a brief transient.

As an attempt to stabilize the temporary capillary network, we
introduce the inelastic repulsion force $\mb F_\mathrm{IR}$, intended
to glue together colliding cells of certain effective radius (in
co-operation with the actractive chemotactic field) and to reduce
bouncing effects (due to its inelastic character), and the persistence
force $\mb F_\mathrm{T}$, which should provide a better alignment of
cells such to enhance the network-like structure.  The inelastic
repulsion term $\mb F_\mathrm{IR}$, because of the effective cell
radius introduced, imposes a limit on the number of cells that can be
inserted in the simulation box without overlap. Our simulations were
set up with cell densities equal to the experimental values, i.e.\ 
varying from $80$~cells/mm$^2$ to $450$~cells/mm$^2$.

Despite our intention of improving the model, we find there is no
clear formation of capillary networks.  By examining
Fig.~\ref{fig:hctr}A and Fig.~\ref{fig:hctr}B, we find disconnected
patterns for $N=80\sim150$~cells, as also foreseen by the experiments,
while we do not find a clear evidence of network formation with
$N=250$~cells, although empty spaces of characteristic area $\ell\,^2$
can be observed in Fig.~\ref{fig:hctr}C and Fig.~\ref{fig:hctr}D. In
particular, the case of $N=450$~cells shows thick structures of cells
reminiscent of the \emph{swiss cheese} pattern, but still disposed in
a rather unorganized geometry.  In Fig.~\ref{fig:net}, we show the
possible resulting network in the case of 250 cells, by drawing cell
connections by hand (by connecting cell neighbors by eye) 
and by superposing a circle with diameter $\ell$.
As in the simpler point-like cell model, the observed (unclear)
network-like structures continue to appear during a short transient,
although with longer lifetimes, after simulation times of
$T\simeq13$~hours, comparable with the experimental ones. Thus, the
introduction of the persistence and inelastic repulsive forces tends
to stabilize the network-like geometry.

The network-like capillary structure raising only during a transient
in the simulation is an unrealistic feature common to the
hydrodynamical models as well \cite{Tor1,gambapc}.  A possible explanation of
this fact, which may lead to more realistic simulations, is that cells
have been modeled as (quasi) static geometrical entities.  In fact, in
reality, cells elongate their shape in the act of moving toward higher
chemo-attractant concentrations \cite{merks}, which is a process
intimately bound to the phenomenon of cell persistence of motion.
This is the main reason that led us to introduce the persistence force
$\mb F_\mathrm{T}$. With this term, in fact, the transient phase gets
longer and the network structures become slightly more clear.  In the
elongated form and after they come in contact, the real cells stick
together, in some sense imitating those connections that we added by
hand in Fig.~\ref{fig:net}.

\section{Conclusions}
\noindent 
We showed that two-dimensional molecular dynamics simulations of
point-like cells whose motion is governed by the process of
chemotaxis, reproduce the main features of real cells to form networks
having chords with a characteristic length. However, this networks are
metastable and collapse after a brief transient.  Similar results are
present in literature, where the problem is faced by solving PDE of
hydrodynamic type, with individual cells represented by gaussian
bumps. To our knowledge our method is the first attempt to describe
endothelial cell systems in terms of a discrete collection of
particles, whose motion can be tracked throughout the whole history of
capillary network formation. The peculiar advantage of molecular
dynamics methods is the extreme ease with which one can introduce
forces acting to individual particles.

In particular, in order to stabilize the metastable network state, we
introduce both an inelastic repulsion force between cells and the so
called ``persistence'' force, which mimics the observed tendency of
cells to preserve the direction of their motion.  Unfortunately, this
model, although more complete and closer to reality, is not fully able
to reproduce the formation of a clear capillary network in
correspondence of realistic cell densities beyond the percolation
threshold. A vague network-like structure appears during a short
transient, whereas the introduction of the persistence of motion force
and the inelastic repulsive force helps to lengthen this metastable
geometry, before all cells collapse in a unique large cluster.
We ascribe this unsuccess to the geometry of single cells, which in
reality change their shape by elongating towards the direction of
motion, a feature that we are not considering at the moment.

\section*{Acknowledgments}
\noindent This research has been partly supported by the TAGora
project funded by the Future and Emerging Technologies program
(IST-FET) of the European Commission under the EU RD contract
IST-34721. 
P.B. and L.T. kindly acknowledges partial support from COFIN-MIUR.


{}

\end{document}